\documentclass{epl}

\title{Persistence of a pinch in a pipe}
\author{L. Mahadevan\inst{1}\thanks{Email: \email{lm@deas.harvard.edu}} \and A. Vaziri \inst{1} \and Moumita Das\inst{1}}

\institute{                    
  \inst{1} Division of Engineering and Applied Sciences, Harvard University, Pierce Hall, 29 Oxford Street, Cambridge MA 02138, USA. \\
}

\begin{document}

\maketitle

\begin{abstract}
The response of low-dimensional solid objects combines geometry and physics in unusual ways, exemplified in structures of great utility such as a thin-walled tube that is ubiquitous in nature and technology.   Here we provide a particularly surprising consequence of this confluence of geometry and physics in tubular structures: the anomalously large persistence of a localized pinch in an elastic pipe whose effect decays very slowly as an oscillatory exponential with a persistence length that diverges as the thickness of the tube vanishes, which we confirm experimentally. The result is more a consequence of geometry than material properties, and is thus equally applicable to carbon nanotubes as it is to oil pipelines.
\end{abstract}


\section{Introduction and Formulation}

\begin{figure}
\onefigure[width=10cm]{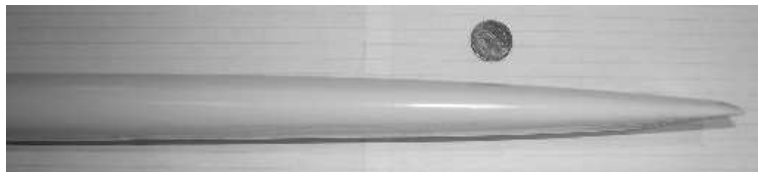} 
\caption{A photograph of a tube pinched at an end shows that the effects of the pinch persist on scales that are much larger than the diameter of the tube. The penny in the background has a diameter of about $1 cm$}
\label{fig1}
\end{figure}

Cylindrical structures are ubiquitous in nature and technology over a range of length scales, from carbon nanotubes and cytoskeletal microtubules to aircraft fuselages. The separation of geometric scales inherent in these structures gives them their large specific stiffness (per unit weight) and endows them with strongly geometrical modes of deformation that involve faceting and other localized modes when subject to pressure, compression, twisting and bending. Here we focus on a peculiar global mode of deformation that is also ubiquitous. Our starting point is the simple observation that when a paper or plastic drinking straw of length $L$, radius $R$ and thickness $t$  ($L \gg R \gg t$) is pinched at an end so that it becomes elliptical locally, as shown in Figure \ref{fig1} (try it yourself); the deformation of the straw persists over a length that is much larger than the radius of the straw. This raises a natural question: what is the persistence length of a pinch ? 

To analyze this, we start with the equations of equilibrium for a shallow cylindrical shell, the von Karman- Donnell equations \cite{b.donnel}:
  \begin{eqnarray}
 \label{vonkarmaneqn}
 B \Delta^2 w- \frac{1}{R} \phi_{,xx}&=& w_{,xx} \phi_{,yy} + w_{,yy} \phi_{,xx}- 2 w_{,xy} \phi_{,xy} \nonumber \\
 \Delta^2 \phi + \frac{E t}{R} w_{,xx} &=& Et ({w_{xy}}^2 - w_{,xx} w_{,yy})
 \end{eqnarray}
Here $A_{,b}=\frac{\partial A}{\partial b}$, $w(x,y)$ is the deflection of the cylinder relative to its naturally curved state parametrized in terms of the azimuthal coordinate $y$ and the axial coordinate $x$, $\phi(x,y)$  is the Airy stress function whose derivatives are the components of the in-plane stress tensor,  $A_{,xx} \equiv \partial^2{A}/{\partial x}^2$ etc., and  $B=Et^3/ 12(1-\sigma^2)$ is the bending stiffness of the sheet of material with Young's modulus $E$ and Poisson's ratio  $\sigma$. The first equation quantifies the balance of forces perpendicular to the cylinder surface, while the second is a geometric compatibility relation involving the in-plane strains. We note that the Karman-Donnell system is one of the simplest of a class of approximate equations of increasing sophistication for the deformations of elastic shells \cite{Libai}; however they are sufficient to explain the phenomena at hand as we shall see.

\section{Analysis, Simulation and Experiment}

When one end of the cylinder is pinched locally exciting the first elliptical mode in the azimuthal direction, we assume that the axial variations induced are on length scales large compared to the radius. Thus suggests a solution of the form $w(x,y)= W(x) \sin(\pi y/R); \phi(x,y)=\Phi(x) \sin(\pi y/R)$. Substituting into Eq. (\ref{vonkarmaneqn}), and keeping only the leading order terms, we find that 
\begin{eqnarray}
(\frac{\pi}{R})^4BW-\frac{1}{R}\Phi'' &\approx& 0 \nonumber\\
(\frac{\pi}{R})^4 \Phi+\frac{Et}{R} W''&\approx 0
\end{eqnarray}
where $(.)'= d(.)/dx$, and we have dropped the nonlinear terms in (\ref{vonkarmaneqn}). Eliminating the function $\Phi(x)$ from the above equations, we find that the amplitude of the deformation $W(x)$ satisfies the linear equation
 \begin{equation}
\label{Wfinaleqn}
W^{''''} + \frac{t^2}{12(1-\sigma^2)} \frac{\pi^8}{R^6} W \approx 0
\end{equation}
It is useful at this point to recall that the balance of forces on an elastic beam supported on an elastic foundation of stiffness $K$ leads to a fourth order equation \cite{Landau} $B W'''' +KW = 0$. Indeed, for axisymmetric deformations of a cylindrical shell, $K = Et/R^2$, so that $Et^2/(12(1-\sigma^2))W''''+ W/R^2 =0$, which while superficially similar to (\ref{Wfinaleqn}) is qualitatively
different; in the limit $t \rightarrow 0$, the former equation is of a singularly perturbed type exhibiting boundary-layer like regions where the solution changes rapidly, quite unlike (\ref{Wfinaleqn}). Indeed, we can see this immediately by noting that the general solution to (\ref{Wfinaleqn}) is
\begin{equation}
W(x)=W_0 \exp(-kx) \cos(kx+\alpha), ~~~k= \left[ \frac{t^2 \pi^8}{12 (1-\sigma^2) R^6}\right ]^{1/4}. \label{scaling}
\end{equation}
where the amplitude $W_0$ and the phase $\alpha$ are determined by the boundary conditions at the end where the pipe is pinched. Thus an applied pinch of amplitude $W_0$ decays as an oscillatory exponential with a characteristic persistence length  $\ell_p=2\pi/k \sim1.1 R^{3/2}/t^{1/2}$, for typical materials  ($\sigma \in [0.33,0,5]$), a scaling hinted at in \cite{b.pablo} with no derivation. The ellipticity induced by the pinch rotates slowly as one progresses along the pipe, analogous to the polarization of a wave.  Since $\ell_p/R \sim (R/t)^{1/2}$, our assumption that the shape and stress have  variations with gradients that are smaller in the axial direction than in the azimuthal direction is justified. We also observe that the persistence length is determined primarily by the geometry of the tube; the dependence on the material properties via  Poisson's ratio is very weak since  $\sigma \in [-1,1/2]$  for isotropic materials. Even more surprising is the anomalous and counter-intuitive divergence of the persistence length as the thickness of the tube vanishes or as the radius of the cylinder diverges, i.e. as the tube is flattened. This geometric amplification is a result in sharp contrast with the focusing that leads to the generation of fine scales and singularities in the inhomogeneous deformation of thin films in such instances as wrinkling and crumpling.
 
\begin{figure}
\onefigure[width=10cm]{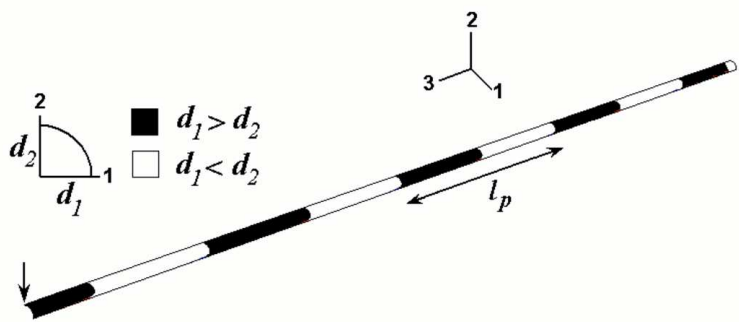}
\caption{Numerical simulations of pinching a cylindrical pipe ($t/R=0.01$) show that the response of the cylinder is indeed oscillatory. The black regions have the same ellipticity as the pinch at one end with $d_1>d_2$, and white regions have an ellipticity that is rotated by $90^{\circ}$ so that $d_1 <d_2$. } 
\label{fig2}
\end{figure}

\begin{figure}
\onefigure[width=12cm]{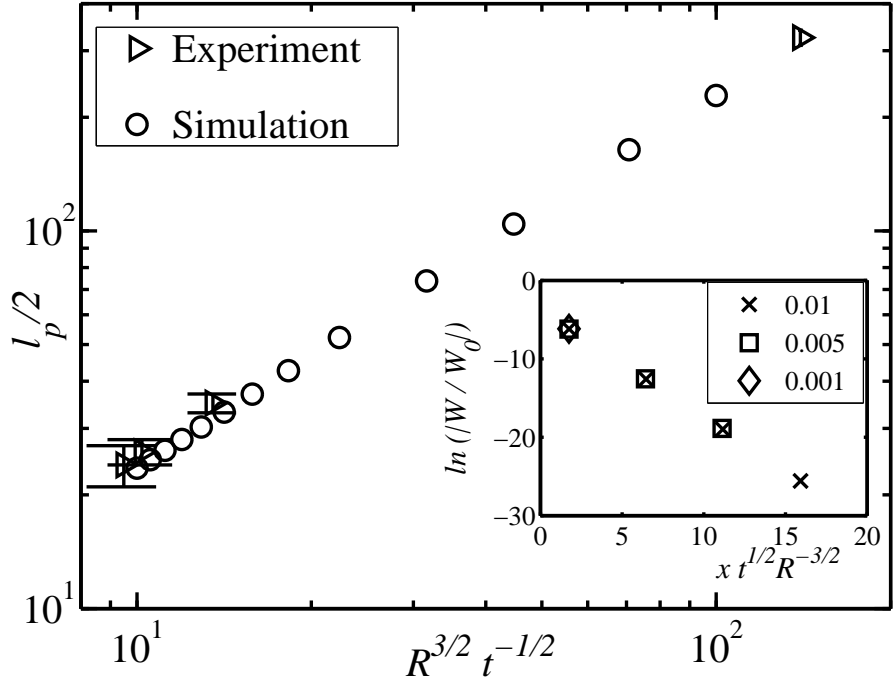}
\caption{Main - Numerical simulations (circles) show that the persistence length of a pinch $\ell_p=C R^{3/2}/ t^{1/2}$, with $C \approx4$, consistent with experiments (triangle),  and the solution of  Eq. (\ref{Wfinaleqn}). The discrepancy in the prefactor in the analysis arises from the linearized analysis of the approximate nature of the shallow shell theory in Eq.(\ref{vonkarmaneqn}). The bars show the standard deviation of the experimental observations. Inset - we show the exponential decrease in the amplitude of the oscillatory response $W(x)$  as a function of the scaled distance $x R^{-3/2}t^{1/2}$ from the applied pinch, obtained from numerical simulations for finite deformations, in agreement with the solution of Eq.(\ref{Wfinaleqn}). The symbols refers to the different values of $t/R$ used in the simulations. }
\label{fig3}
\end{figure}

The above linearized analysis is valid only for small deformations in light of the approximate nature of the von-Karman-Donnell equations and our asymptotic analysis. To check the validity of our analysis, we carried out numerical simulations using the finite element method. The pinch was applied incrementally and a commercial package ABAQUS was used to minimize the elastic energy of a linear isotropic solid in a slender geometry. For very thin shells, this elastic energy density that is minimized is approximately:
\begin{eqnarray}
 U&=& \frac{Et}{2(1-\sigma^2)} \left[({\epsilon_1}+ {\epsilon_2})^2 - 2 (1-\sigma )(\epsilon_1 \epsilon_2 - \gamma^2) \right] \nonumber \\
 &+&\frac{Et^3}{24(1-\sigma^2)} \left[({\kappa_1}+ {\kappa_2})^2 - 2 (1-\sigma)( \kappa_1 \kappa_2 - \tau^2) \right].
 \end{eqnarray}
Here  $\epsilon_1, \epsilon_2, \gamma$  are the in-plane extensional and shear strains and $\kappa_1, \kappa_2, \tau$   are the curvatures and twist relative to the undeformed state of the tube. The first term in the expression  accounts for in-plane deformations and the second term accounts for out-of-plane deformations. Four-node, quadrilateral shell elements with reduced integration and a large-deformation formulation to account for the finite curvatures of the shell were used in calculations. A sensitivity study was conducted to ensure the independence of the results on the computational mesh. We considered very thin isotropic and homogeneous elastic tubes which were pinched symmetrically at one end with a prescribed indentation deformation $w(0,0)=-w(0,\pi R)=W_0$. The tubes were assumed to be made  of linearly elastic material (with Young's modulus $E = 100$ MPa and Poisson's ratio $\sigma = 0.3$ ) of varying $t$ in our simulations, with $t \ll R \ll L$. In Figure  \ref{fig2} we see that the tube responds with an oscillatory deformation of varying ellipticity consistent with our simple analytical predictions. In Fig., \ref{fig3}, we show that the  persistence length measured numerically by following the period of oscillation of the ellipticity also follows our analytical predictions; the numerical experiments were carried out for tubes of various $ t$ keeping the applied pinch amplitude $W_0$ and the tube radius $R$ constant. In the inset we show that the amplitude of the deformation of the tube decays exponentially away from the location of the pinch. However, there is a discrepancy between the prefactor of the wavelength determined numerically and that in (\ref{scaling}) which may be attributed to the approximate nature of the von Karman-Donnell equations and our linearized analysis thereof.

We also carried out experiments with long straws and tubes of various thickness (ranging from 0.1 mm to 0.3 mm) and radii (ranging from 3 mm to 12.5 mm). The images of the pinched straws were taken with a digital camera and were analyzed using the  Canny edge-detection-algorithm in MATLAB to extract the information on boundaries of the deformed profile. The thickness and radius of the tubes were measured using slide calipers. In Fig. \ref{fig3}, we show that the  experimentally measured values of the persistence length match the numerical simulations  and are consistent with the scaling derived analytically. 


\section{Discussion}

Before we discuss our results, it is worthwhile to compare and contrast our results for the persistence of a pinch in cylinder with that of a pinch in a narrow flat plate of width $w$ and length $L ( \gg w)$ (equivalently, we could consider periodically pinched sheet such as a drape). When such a sheet is pinched at an end to make it slightly narrower so that the amplitude of the pinch is $a$, casual observations show that the persistence of the pinch is much larger than the width of the strip. To understand this persistence, we note that the dominant component of the curvature scales as as $a/w^2$ so that bending energy in the sheet is roughly $Et^3(a/w^2)^2 wl_p$, where $l_p$ is the persistence length of a pinch. Over this length scale,  the sheet is weakly curved in two directions so that it must be stretched (since the Gauss curvature of a doubly curved surface is non-zero, it follows from Gauss's theorema egregium that it must have been stretched), with a typical strain that scales as $a^2/l_p^2$, costing an energy that scales as $Et(a^2/l^2)^2l_pw$. Minimizing the sum of the bending and stretching energy, we see that the persistence length of a pinch scales as $l_p \sim w (a/t)^{1/2}$ \cite{Lobkovksy97, Cerda04}, showing a clear dependence on the amplitude of the pinch. However, for a pinched pipe, when $a \sim w \sim R$, we find that $l_p \sim R^{3/2}/t^{1/2}$ consistent with the result obtained here.

Some  immediate implications follow from our results: in molecular  carbon nanotubes \cite{b.harris,b.srivastava} or cytoskeletal microtubules \cite{b.pablo,b.howard} our result suggest how mechanical strains may be transmitted over long distances along tubes. Since the energy associated with pinching the pipe scales as $Et^{5/2} R^{1/2}$, the stiffness associated with a pinching deformation scales as $Et^{5/2}/R^{3/2}$ is quite small, and raises the question of whether nature might already use this mode. On mesoscales, a dramatic example of the failure of tubes is afforded by the flip-flop buckle propagation in submarine pipelines \cite{b.palmer}, wherein a pipe flattens for a while in one direction and then in an orthogonal direction, with a characteristic length scale of the flattened regions. Our result complements numerical simulations of the process \cite{b.park} by providing a simple explanation for the flip-flop which we interpret as a simple consequence of a globally deformed tubular shell  induced for example by a local pinch. Then, an oscillatory elliptic mode follows naturally, and sets the stage for dynamic and plastic buckling that will likely lead to a flip-flop mode of propagation. Thus our calculation immediately suggests a criterion for the design of buckle arrestors - which might be spaced at a distance comparable to $\ell_p$. Finally, on planetary length scales, the curved nature of continental and oceanic plates together with our calculation suggests a possible mechanism for the very long scale persistence of deformations induced by the curvature of the crust that would otherwise sink far more quickly under its own weight \cite{b.kearey}.

\end{document}